\documentstyle[11pt,paspconf,psfig]{article}

\let\simgt\ga
\let\simlt\la
\def\kms{\hbox{km\,s$^{-1}$}}
\begin{document}

\vbox to0pt{\vskip-0.7cm
\centerline{\it ~~~~To appear in Proc.\ IAU Coll.\ 160, ``Pulsars:
Problems and Progress''}\vss}

\title{The White Dwarf Companions of Recycled Pulsars}

\author{M. H. van Kerkwijk}
\affil{Palomar Observatory, California Institute of Technology 105-24,
Pasadena, CA 91125, USA}

\begin{abstract}
I review what properties of the white-dwarf companions of recycled
pulsars can be inferred from optical observations, and discuss how
these can help us understand the characteristics and evolution of
these binaries.  I focus on spectroscopic observations, describing
results obtained recently, and looking forward to what may come.
\end{abstract}


\section{Introduction}

Our understanding of the characteristics and evolution of neutron-star
binaries would benefit greatly from a better grasp of the properties
of the companions.  For instance, for recycled pulsars with low-mass
helium white dwarf companions (low-mass binary pulsars or LMBP),
evolutionary theory (for a review, see Bhattacharya \& Van den Heuvel
\cite{bhatvdh:91}) predicts a relation between the orbital period and
the companion mass (Rappaport et al.\ \cite{rapp&a:95} and references
therein).  So far, it has only been possible to verify this prediction
for one system, PSR~B1855+09, for which Kaspi et al.\
(\cite{kasptr:94}) determined the companion mass from Shapiro delay in
the pulse arrival times.

Another prediction from evolutionary theory is that the neutron star
in a LMBP should have accreted up to $0.7\;M_\odot$ (e.g., Van den
Heuvel \& Bitzaraki \cite{vdheb:95}).  Thus, if the neutron star
started with the ``canonical'' $1.4\;M_\odot$ (for a recent census,
see Van Kerkwijk et al.\ \cite{vkervpz:95}), it would now be
$\simgt\!2\;M_\odot$.  This would strongly constrain the equation of
state (EOS) at supranuclear densities (e.g., Cook et al.\
\cite{cookst:94}), since for softer EOS, like the one recently
proposed by Brown \& Bethe (\cite{browb:94}), such a massive neutron
star would collapse into a black hole.  The only LMBP for which a
neutron-star mass estimate is available, is (again) PSR~B1855+09, for
which Kaspi et al.\ (\cite{kasptr:94}) found
$M_{\rm{}NS}=1.50^{+0.26}_{-0.14}\;M_\odot$ (68\% confidence).  As
yet, the uncertainty is too high to allow one to draw a strong
conclusion.

It would be of obvious interest to use direct observations of the
white-dwarf companions to determine properties such as their mass.  So
far, due to their faintness, the observations have mostly been
restricted to photometry, which allows one only to derive a
temperature (by comparison with white-dwarf models, using an estimate
of the surface gravity).  The temperature can be used to constrain the
cooling age, and such a constraint has led to one of the first
indications that magnetic fields of pulsars do not decay (Kulkarni
\cite{kulk:86}).  For an accurate measure, however, one also needs
information about the mass, radius, and internal composition of the
white dwarf.

As discussed in Sect.~\ref{sec:wd}, spectroscopy of the white-dwarf
companions can be used to determine masses and radii, and thus may
allow one to verify the orbital-period/mass relation, and to obtain
better limits on the cooling age.  Furthermore, for the shorter
orbital periods, one can measure the radial-velocity amplitude and
determine the mass ratio.  Combined with the white-dwarf mass, this
determines the mass of the neutron star.  

Given these possibilities, it is clearly worthwhile to try to obtain
spectra of the white-dwarf companions.  Unfortunately, there are not
many that are bright enough, and before the $10\;$m Keck telescope
became available, only the companion of PSR~J0437$-$4715 was observed
spectroscopically (Danziger et al.\ \cite{danzbdv:93}).  This white
dwarf, however, was too cool to show any features (see
Sect.~\ref{sec:indiv}).  Recently, we have obtained a number of
spectra of other white-dwarf companions (Van Kerkwijk \& Kulkarni
\cite{vkerk:95}, hereafter Paper~I; Van Kerkwijk et al.\
\cite{vkerbk:96}, hereafter Paper~II), which all show more interesting
spectra (Sect.~\ref{sec:indiv}).

\section{Measuring a White Dwarf\label{sec:wd}}

A number of methods have been employed to determine fundamental
parameters of white dwarfs, such as their mass, radius, and
luminosity.  In general, if one can measure any combination of the
mass and radius, one can determine the mass and radius separately
using the mass-radius relation.  This relation is reasonably well
understood, since white dwarfs have a relatively simple structure,
which, for low enough temperatures, is determined by the degeneracy of
the electrons (Chandrasekhar \cite{chan:39}; Hamada \& Salpeter
\cite{hamas:61} for a detailed study).  One can do better by using
evolutionary models in which finite-temperature effects and the
thickness of the surface hydrogen and/or helium layers are taken into
account (see Bergeron et al.\ \cite{bergsl:92}).  Such models have
been produced by, e.g., Wood (\cite{wood:92}, \cite{wood:95}).  Having
determined the mass and radius, the same models can be used to
determine the age of the white dwarf from the temperature.

A number of different methods have been used to infer white-dwarf
masses (discussed in some detail by Bergeron et al.\
[\cite{bergsl:92}]).  They are based on determinations of: (i) the
radius, from accurate distance, flux and temperature estimates; (ii)
the gravitational redshift, using spectroscopically determined
velocity shifts between lines from the white dwarf and a reference,
usually a common-proper-motion companion
($GM/Rc=0.635(M/M_\odot)(R/R_\odot)^{-1}\;\kms$); (iii) the surface
gravity, determined from fitting the line profiles to model
atmospheres; (iv) mode analysis of pulsating white dwarfs; and (v) the
orbit for a white dwarf in a resolved binary.  Of course, to this list
one should add: (vi) Shapiro delay for white-dwarf, pulsar binaries in
nearly edge-on orbits.

Of these, only the one using the surface gravity can be applied to a
large number of objects.  The model-atmosphere calculations required
for this purpose are relatively straightforward, as usually only one
element has to be taken into account---the elements having been
separated due to the high gravity---and as the pressure is high enough
for LTE to be a good approximation.  The technique is furthest
developed for the white dwarfs with hydrogen atmospheres (called `DA';
for a review, Wesemael et al.\ \cite{wese&a:93}), and masses for a
large number of DA white dwarfs have been determined (e.g., Bergeron
et al.\ \cite{bergsl:92}).  

%Recently, accurate model atmospheres for objects showing helium in
%their spectra (the `DB' stars) have become available as well
%(Beauchamp et al.\ \cite{beau&a:96}).

There are a two possible problems in applying this method to the
white-dwarf companions of recycled pulsars, both related to the fact
that they are, in general, less massive and cooler than the white
dwarfs studied so far.  The lower mass implies that they have helium
in their core, instead of the usual carbon and oxygen, and for these
helium white dwarfs only the Hamada-Salpeter zero-temperature
mass-radius relation is available.  This situation should improve,
however, as cooling models specifically for helium white dwarfs are
being produced (Hansen \& Phinney \cite{hansp:96}; Phinney, these
proceedings).

A problem related to the lower temperatures is that at
$T\simlt12000\;$K the hydrogen layer becomes convective, and helium
from the core might get dredged up and mixed in.  The additional
pressure exerted by the helium would be very hard to distinguish from
an increased surface gravity (Bergeron et al.\ \cite{bergwf:91}),
except perhaps in the infrared (Bergeron et al.\ \cite{bergsw:95}).
Thus, ignoring it would lead to an overestimate of the mass.  Whether
or not this will be a problem, depends mostly on whether the hydrogen
layer is thin enough for the convection zone to reach the helium core.
The best constraints, derived from asteroseismology of pulsating white
dwarfs (e.g., Fontaine et al.\ \cite{font&a:94} and references
therein), indicate rather thick hydrogen layers
($\sim\!10^{-4}\;M_{\rm{}WD}$), consistent with predictions from
evolutionary models (for a review, D'Antona \& Mazzitelli
\cite{dantm:90}).  For such thick layers, little mixing should occur.
From a comparison of spectroscopic masses with those derived from
gravitational redshifts, however, it seems that for cooler
temperatures the former are systematically higher than the latter,
indicating that helium may in fact be present (Reid \cite{reid:96}).

\section{Results for Individual Systems\label{sec:indiv}}

\subsubsection{PSR~J0437$-$4715} ($P=5.76\;$ms,
$P_{\rm{}orb}=5.74\;$d; for the most recent timing results, see Bell
et al.\ \cite{bell&a:95}) is the closest recycled pulsar known.  A
candidate companion was identified on the UK Schmidt sky survey by
Johnston et al.\ (\cite{john&a:93}).  This identification was
confirmed in follow-up studies by Bailyn (\cite{bail:93}), Bell et
al.\ (\cite{bellbb:93}), and Danziger et al.\ (\cite{danzbdv:93}).
These authors found that the white dwarf is relatively bright and red
($V=20.9$, $B-V=1.3$ [note that the photometry listed in the three
papers is not quite consistent]).  The inferred temperature is about
$4000\;$K.  Danziger et al.\ (\cite{danzbdv:93}) found that the
spectrum is featureless.  This is not unexpected, since in white
dwarfs both hydrogen and helium are spectroscopically invisible at
$\sim\!4000\;$K.

It may be possible to constrain the surface gravity (and the
abundances) somewhat from accurate optical and near-infrared
photometry (see Bergeron et al.\ \cite{bergsw:95}).  For this
relatively nearby pulsar, however, a better handle on the mass can
probably be obtained by determining an accurate distance (to better
than 10\% or so).  For this system, there are four possibilities: (i)
VLBI parallax; (ii) timing parallax; (iii) the proper motion combined
with the velocity inferred from optical observations of the bow shock
(as has been done for PSR~B1957+20 by Aldcroft et al.\
\cite{aldcrc:92}); and (iv) the proper motion combined with the
apparent decay of the orbital period (Bell \& Bailes \cite{bellb:96};
Bell, these proceedings).  The distance combined with the observed
flux and temperature can be used to infer the radius and thus, with
the mass-radius relation, the mass of the white dwarf.

\subsubsection{PSR~B0655+64} ($P=196\;$ms, $P_{\rm{}orb}=1.03\;$d;
Jones \& Lyne \cite{jonel:88}; Taylor \& Dewey \cite{tayld:88}) has,
given the rather high companion mass of $\sim\!0.7\;M_\odot$ indicated
by the mass function, most likely been through a phase as a high-mass
X-ray binary, followed by spiral in (e.g., Bhattacharya \& Van den
Heuvel \cite{bhatvdh:91}).  Presumably, the helium core left was not
massive enough to form a second neutron star.  Such low-mass helium
stars become giants (Paczynski \cite{pacz:71}; Habets \cite{habe:86}),
and a second stage of mass transfer may ensue.  It seems likely that
in this stage any remaining hydrogen will disappear.  Thus, one
expects to be left with a white dwarf with a carbon-oxygen core and a
helium atmosphere.

\begin{figure}
\parbox[b]{0.5\textwidth}{\psfig{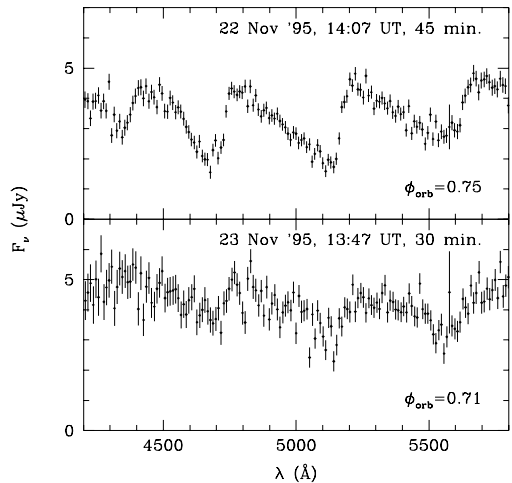}}%
\parbox[b]{0.5\textwidth}{%
\caption[]{Two spectra of the companion of PSR B0655+64, taken at
approximately the same orbital phase.  (Note that the observing
conditions were not photometric.  Hence, the absolute flux levels are
not reliable.) The large change in strength of the Swan C$_2$ bands
confirms the conclusion drawn in Paper~I that the changes are not
related to orbital effects.  From all spectra taken in November 1995,
a periodicity of $9.7\pm0.1\;$h is indicated.\label{fig:b0655}}}
\end{figure}

The companion was identified by Kulkarni (\cite{kulk:86}).  It has
$V=22.2$ and $V-R=0.1$, indicating a temperature of about 6000 to
$9000\;$K.  First spectra were taken on New Year's Eve 1995 with the
Keck telescope (Paper~I).  These spectra showed strong absorption
bands of C$_2$, the so-called Swan bands (see also
Fig.~\ref{fig:b0655}).  White dwarfs showing carbon are called DQ
stars, and have for a long time posed a major puzzle, since it was not
understood how there could be carbon in the photosphere, when gravity
should long since have separated the elements.  It has become clear
that it is because trace amounts of carbon are dredged up when the
helium atmosphere becomes maximally convective, at $T\simeq12000\;$K
(Pelletier et al.\ \cite{pell&a:86}).  Subsequently, the carbon will
only slowly be depleted from the convective helium layer.  Thus, the
detection of carbon in the companion directly confirms that it is a
carbon-oxygen white dwarf.

Unlike what is seen in other DQ stars, the Swan bands were seen to
vary in strength, by about a factor two in less than two hours
(Paper~I).  This seems to indicate that the white dwarf has brighter
and darker spots on its surface, possibly due to a magnetic fields (a
locally higher magnetic field strength would lead to a lower
temperature---like in a sunspot---and thus to much stronger
molecular bands).  The change in strength was shown to be too fast and
too large to be due to orbital modulation.  This is confirmed by
additional spectra obtain in November 1995, which show widely
different band strengths at the same orbital phase
(Fig.~\ref{fig:b0655}).  Most likely, the modulation reflects the
white-dwarf rotation.  From the November data, a period of
$9.7\pm0.1\;$h is indicated.

In Paper~I, it was noted that if the helium-giant progenitor was
filling its Roche lobe and rotating synchronously, it would have been
spun up due to conservation of angular momentum when it shrunk to form
a white dwarf.  If the spin-up is mostly due to the angular momentum
contained in the remaining giant envelope, the envelope mass must have
been $\sim\!10^{-4.5}\;M_\odot$.  Interestingly, this is similar to
the helium-layer masses that are inferred using other techniques
(Pelletier et al.\ \cite{pell&a:86}; Weidemann \& Koester
\cite{weidk:95}; Dehner \& Kawaler \cite{dehnk:95}).

\subsubsection{PSR~B0820+02} ($P=865\;$ms, $P_{\rm{}orb}=1232\;$d; Taylor
\& Dewey \cite{tayld:88}) has the longest orbital period of all LMBPs
known.  Thus, it provides an excellent opportunity to test the
orbital-period, mass relation.  Its companion was identified by
Kulkarni (\cite{kulk:86}) and studied in more detail by Koester et
al.\ (\cite{koescr:92}).  It is bright enough ($V=22.8$) for
spectroscopy.  A first spectrum shows that it is a DA white dwarf
(Paper~I), and thus a mass determination via the surface gravity will
be possible (fortunately, the temperature of 14000--16500$\;$K
[Koester et al.\ \cite{koescr:92}] is high enough that helium will
certainly not be present.).  With about one night of Keck
observations, we hope to be able to do this.

\subsubsection{PSR~J1012+5307} ($P=5.26\;$ms, $P_{\rm{}orb}=0.60\;$d;
Lorimer et al.\ \cite{lori&a:95}) has a relatively bright optical
counterpart ($V=19.6$), found on the Palomar sky survey by Nicastro et
al.\ (\cite{nica&a:95}).  Astrometry and photometry by Lorimer et al.\
(\cite{lori&a:95}) confirmed the association, and showed that the
white dwarf was rather hot, $T_{\rm{}BB}=9400\pm300\;$K.  This
indicates a cooling age of a couple $10^8\;$yr, much shorter than the
characteristic age of $7\,10^9\;$yr (but see Alberts et al.\
\cite{albe&a:96}).

\begin{figure}
\centerline{\hbox{\psfig{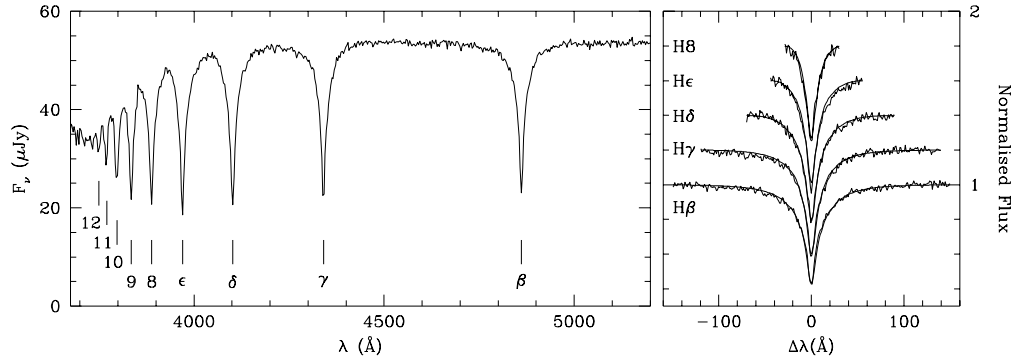}}}
\caption[]{The spectrum of the white-dwarf companion of PSR J1012+5307
(taken from Paper~II), with the Balmer lines indicated.  In the
right-hand panel, the normalized profiles of H$\beta$ to H8 are shown,
with the best-fit model-atmosphere profiles superposed.  The fit gives
$T_{\rm{}eff}=8550\pm25\;$K, and $\log{}g=6.75\pm0.07$ (cgs
units).\label{fig:j1012model}}
\end{figure}

\begin{figure}
\parbox[b]{0.5\textwidth}{\psfig{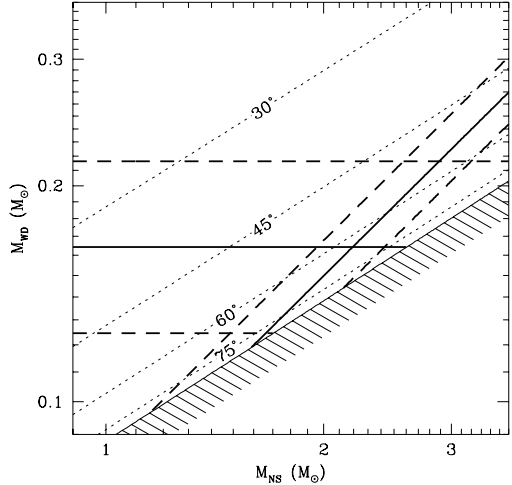}}%
\parbox[b]{0.5\textwidth}{%
\caption[]{The masses of PSR~J1012+5307 and its companion (taken from
Paper~II).  Shown are the constraints from the pulsar mass function
(thin, drawn curve for $i=90^\circ$; dotted curves for inclinations as
indicated), the ratio of the radial-velocity amplitudes (thick,
slanted curve), and the white-dwarf surface gravity (thick, horizontal
curve).  For the latter two, the 95\% uncertainty regions are
indicated by thick, dashed curves.\label{fig:j1012masses}}}
\end{figure}

The spectrum of the companion of PSR~J1012+5307 shows hydrogen lines,
from H$\alpha$ up to H12 (Paper~II; see Fig.~\ref{fig:j1012model}).
The presence of the higher Balmer lines indicates that the object is a
low-mass white dwarf.  From a model-atmosphere fit
(Fig.~\ref{fig:j1012model}), a temperature $T_{\rm{}eff}=8550\pm25\;$K
and a surface gravity $\log{}g=6.75\pm0.07$ (cgs units) are derived.
Using the Hamada-Salpeter zero-temperature relation, with an
approximate finite-temperature correction, we infer a mass
$M_{\rm{}WD}=0.16\pm0.02\;M_\odot$ (the lowest among all
spectroscopically identified white dwarfs).

We also derive radial velocities, and find a radial-velocity amplitude
$K_{\rm{}WD}=280\pm15\;\kms$ and systemic velocity
$\gamma=-50\pm15\;\kms$.  The implied mass ratio is
$M_{\rm{}NS}/M_{\rm{}WD}=13.3\pm0.7$.  From this mass ratio, the mass
of the white dwarf, and the pulsar mass function, we find that with
95\% confidence $1.5<M_{\rm{}NS}/M_\odot<3.2$ (see
Fig.~\ref{fig:j1012masses}).

This determination is not yet accurate enough to constrain the
equation of state, or to test evolutionary theory, but it does show
that further study may well prove fruitful.  It will be relatively
straightforward to improve the accuracy of the mass ratio, which, as
is clear from Fig.~\ref{fig:j1012masses}, might already lead to an
interesting constraint on the mass of the neutron star.  It will be
less easy to improve the white-dwarf mass estimate.  This is because
of the uncertainties in the mass-radius relation for these very
low-mass white dwarfs, and because of the possible presence of helium
in the atmosphere.  As discussed above, if helium is present, the true
surface gravity---and thus the inferred mass---will be lower.

The pulsar is relatively nearby and bright, however, and it may well
be possible to derive an accurate distance.  This would allow one to
obtain a direct estimate of the radius.  If this is the same as the
predicted one ($0.028\pm0.002\;R_\odot$), it would give confidence in
the result.  If it is not, one can either assume there is a problem
with the mass-radius relation, but not with helium pollution, and
infer a mass from the radius in combination with the observed surface
gravity, or one can assume that there is helium pollution, but that
the mass-radius relation is fine, and use that to derive a mass from
the radius.  Another possibility is to search carefully for Shapiro
delay in the pulse arrival times.  As one can see from
Fig.~\ref{fig:j1012masses}, the inclination should be
$\simgt\!60^\circ$ for $M_{\rm{}NS}\simlt2\;M_\odot$.

\section{Plans and Conclusions\label{sec:plans}}

Two other white-dwarf companions, of PSR~J1022+1001 (Camilo
\cite{cami:95}) and of PSR~J0218+4232 (Navarro et al.\
\cite{nava&a:95}), are bright enough for spectroscopy.  For the
former, the mass function indicates a massive white dwarf, like for
PSR~B0655+64, and one might hope to find a similarly interesting
spectrum.  We have identified the companion using the $5\;$m Hale
telescope (see also Lundgren, these proceedings), and it is of similar
brightness and temperature as PSR~B0655+64.  If Swan bands are present
too, and our ideas about the cause of the variations are correct, any
periodicity should be faster than in PSR~B0655+64.  For
PSR~J0218+4232, we have found an optical counterpart on Keck images.
With an orbital period of 2 days, it may be possible to do a
radial-velocity study like for PSR~J1012+5307.

A conclusion could be that from binary pulsars one still can hope to
learn sometimes not quite as much as expected, sometimes more, and
sometimes the unexpected.  Combining radio observations of the pulsar
with optical observations the white dwarf, one may hope to
overdetermine the system, so that the consistency of the results can
be verified (like from timing alone for the double neutron-star
binaries).  At present, perhaps the most promising system is
PSR~J1012+5307, for which we found tantalizing indications for a
neutron star that is truly more massive than $1.4\;M_\odot$.  It seems
worthwhile to do an in-depth study of this system, using further
optical radial-velocity measurements to refine the mass ratio, radio
VLBI and timing to determine the distance, and searches for Shapiro
delay to constrain the companion mass and inclination.  Higher-order
effects in the timing might help constrain the inclination, as could
perhaps the orbital variation of the scintillation velocity (see Lyne
\cite{lyne:84}).  For this system, as well as the others, the coming
years should prove interesting.

\acknowledgements I thank Brad Hansen, Shri Kulkarni, Yanqin Wu,
Pierre Bergeron, Fernando Camilo, and Andrew Lyne for useful
discussions, and acknowledge support from a NASA Hubble Fellowship.

\end{document}